\documentclass[conference]{IEEEtran}

\usepackage{graphicx}
\graphicspath{{figures/}}
\DeclareGraphicsExtensions{.jpg,.png}

\usepackage{bibspacing}
\usepackage[caption=false,font=footnotesize]{subfig}

\usepackage{amsmath}
\usepackage{stmaryrd}

\usepackage{array}
\usepackage{color}
\usepackage[hyphens]{url}
\usepackage[pdftitle=A\ First\ Look\ at\ the\ Usability\ of\ Bitcoin\ Key\ Management,pdfauthor=Shayan\ Eskandari\ et\ al.]{hyperref}
\usepackage{comment}
\usepackage{cite}

\usepackage{xspace}
\newcommand{\etal}{\textit{et al.}\xspace}
\newcommand{\etc}{\textit{etc.}\xspace}
\newcommand{\ie}{\textit{i.e.,}\xspace}
\newcommand{\eg}{\textit{e.g.,}\xspace}

\renewcommand\paragraph[1]{\smallskip\noindent\textbf{#1.}}

\newenvironment{compactlist}
  {\begin{itemize} 
  \setlength{\itemsep}{0pt} 
  \setlength{\parskip}{0pt}} 
  {\end{itemize}}

\newcommand{\bitcoinclient}{Bitcoin Core\xspace}
\newcommand{\multibit}{MultiBit\xspace}
\newcommand{\armory}{Armory\xspace}
\newcommand{\paper}{Bitaddress\xspace}
\newcommand{\brain}{Brainwallet\xspace}
\newcommand{\block}{Blockchain.info\xspace}
\newcommand{\coinbase}{\texttt{Coinbase.com}\xspace}

\usepackage{adjustbox}
\newcommand{\headrow}[1]{\multicolumn{1}{c}{\adjustbox{angle=45,lap=\width-0.5em}{#1}}}

\newcommand{\full}{$\bullet$}
\newcommand{\prt}{$\circ$}

\begin{document}

\title{A First Look at the Usability of Bitcoin Key Management}

\author{\IEEEauthorblockN{Shayan Eskandari\IEEEauthorrefmark{1}, David Barrera\IEEEauthorrefmark{2}, Elizabeth Stobert\IEEEauthorrefmark{3}, and Jeremy Clark\IEEEauthorrefmark{1}} \IEEEauthorblockA{\IEEEauthorrefmark{1}Concordia University, \IEEEauthorrefmark{2}ETH Z{\"u}rich, \IEEEauthorrefmark{3}Carleton University}
}

\IEEEoverridecommandlockouts
\makeatletter\def\@IEEEpubidpullup{9\baselineskip}\makeatother
\IEEEpubid{\parbox{\columnwidth}{Permission to freely reproduce all or part
    of this paper for noncommercial purposes is granted provided that
    copies bear this notice and the full citation on the first
    page. Reproduction for commercial purposes is strictly prohibited
    without the prior written consent of the Internet Society, the
    first-named author (for reproduction of an entire paper only), and
    the author's employer if the paper was prepared within the scope
    of employment.  \\
    USEC '15, 8 February 2015, San Diego, CA, USA\\
    Copyright 2015 Internet Society, ISBN 1-891562-40-1\\
    http://dx.doi.org/10.14722/usec.2015.23015
}
\hspace{\columnsep}\makebox[\columnwidth]{}}

\maketitle



\begin{abstract}
Bitcoin users are directly or indirectly forced to deal with public key cryptography, which has a number of security and usability challenges that differ from the password-based authentication underlying most online banking services. Users must ensure that keys are simultaneously accessible, resistant to digital theft and resilient to loss. In this paper, we contribute an evaluation framework for comparing Bitcoin key management approaches, and conduct a broad usability evaluation of six representative Bitcoin clients. We find that Bitcoin shares many of the fundamental challenges of key management known from other domains, but that Bitcoin may present a unique opportunity to rethink key management for end users.
\end{abstract}





\section{Introduction}
In all of the excitement surrounding Bitcoin~\cite{Nak08}, it is easy to forget that the decentralized currency assumes a solution to the longstanding problem of usable public key cryptography for user authentication. Studies of the usability of key management~\cite{GM05,GMSN+05,SBKH06,GFF06} have shown that there are numerous usability issues that prevent public key cryptography from being effectively leveraged by end users. Managing, controlling, and using cryptographic keys are complex tasks, and no clear solution has been proposed.  

Despite the known complexity in creating and managing cryptographic keys, the Bitcoin network and software clients use such keys extensively for many operations. For example, digital signatures, which require the Bitcoin software to read private keys into memory, are used to assert ownership over a specific set of Bitcoins. Thus, managing the same coins on multiple devices (\eg a desktop and a phone) requires the corresponding private keys to be copied to and made accessible on these devices. 


The consequences of losing exclusive control over an account containing monetary value connects the threat of losing a Bitcoin private key to that of losing an online banking password. However, consumers in many countries are legally protected from any liability of banking credential loss. Furthermore, most bank transactions are traceable and reversible, making it difficult to extract value from stolen banking credentials (most techniques involve a mule~\cite{FH12}). Bitcoin transactions are also traceable, however they are not reversible. Stolen Bitcoins can thus not be centrally or automatically recovered. Bitcoin users typically have no legal protection against loss or theft, and while stolen Bitcoins could be traced as they change ownership,\footnote{Public keys associated with specific Bitcoins are publicly available in the Bitcoin blockchain, but the identities of users who control those keys are not.} several mechanisms exist for laundering Bitcoins and similar digital currencies~\cite{MGGR13,BNMC+14}.

In an effort to address some of the complexities of key management, developers of Bitcoin software have created a variety of innovative technologies ranging from password-derived keys to air-gapped computers to physical printouts of private keys in the form of 2D barcodes. However, since none of these proposals have been evaluated in the Bitcoin context, it remains unclear which techniques have usability advantages.

For Bitcoin to flourish, adoption must expand beyond developers and tech-savvy enthusiasts to novice users. Expansion solidifies the need for a usable, comprehensible approach to Bitcoin. If users cannot safely manage Bitcoin keys, it may result in the users' loss of funds and/or a poor reputation for Bitcoin, both of which could dissuade further user adoption. 

In this paper, we aim to investigate the usability challenges surrounding key management in Bitcoin. To do this, we survey and categorize the most prominent Bitcoin key management proposals. Next we conduct an expert usability inspection technique known as a cognitive walkthrough~\cite{WRLP94} on popular examples of each proposal. Our goal is to identify overarching usability issues as well as advantages of specific proposals, allowing us to propose design recommendations for future Bitcoin clients.



Specifically, the contributions of the paper are as follows:

\begin{compactlist}
\item We perform a broad survey of six Bitcoin key management techniques which cover the vast majority of deployed Bitcoin software. 

\item Using the results from our survey, we propose an evaluation and comparison framework for Bitcoin key management techniques. The framework is based on 10 security, usability and deployability criteria, and enables direct comparison of current and future key management proposals. Using our framework we find that certain properties, such as trust in a central party enable additional beneficial properties. We also find that the disadvantages of certain properties, such as malware protection, outweigh the relative benefits.

\item We perform a cognitive walkthrough of six distinct Bitcoin clients and tools to identify usability issues while performing basic Bitcoin tasks (\eg viewing account balance, sending funds, \etc). We find that the metaphors and abstractions used in the surveyed clients are subject to misinterpretations, and that the clients do not do enough to support their users.

\end{compactlist}

\section{Background}
\subsection{Bitcoin}
Bitcoin is a cryptographic currency deployed in 2009~\cite{Nak08} which has reached a level of adoption unrealized by decades of previously proposed digital currencies (from 1982~\cite{Cha82} onward). Unlike many previous proposals, Bitcoin does not distribute digital monetary units to users. Instead, a public ledger maintains a list of every transaction\footnote{Technically, a transaction specifies a short script that encodes how the balance can be claimed as the input to some future transaction.} made by all Bitcoin users since the creation of the currency. A \textit{transaction} in its simplest form describes the movement of some balance of the Bitcoin currency (XBT or BTC) from one or more accounts (called input addresses) into one or more accounts (called output addresses). Bitcoin addresses are indexed by the fingerprint of a public key from a digital signature scheme.\footnote{Elliptic Curve Digital Signature Algorithm (ECDSA)~\cite{ecdsa}.} They are not centrally allocated or registered in any way---the addresses become active when the first transaction moving money into them is added to the ledger. 

In Bitcoin, every transaction must be digitally signed using the private signing key associated with each input address in the transaction. In order to spend Bitcoin, users require access to the signing key of the account holding their Bitcoin. Thus users do not maintain any kind of units of currency; they maintain a set of keys that provide them signing authority over certain accounts recorded in the ledger. 

The ledger (known as the \emph{blockchain}) is maintained and updated by a decentralized network using a novel method to reach consensus that involves incentivizing nodes in the network with the ability to generate (known as mining) new Bitcoin and collect transaction fees. The details of the Bitcoin consensus model are not relevant to this paper, but we note that clients in the network participate in the consensus model by downloading and cryptographically verifying the integrity of the blockchain. As of writing, the Bitcoin blockchain is roughly 25 GB in size.\footnote{Due to the large size of the blockchain, full download is infeasible for thin clients running on mobile devices, as well as some desktop clients. These clients connect to a semi-trusted node and only request transactions relevant to keys in their wallet. This technique, known as Simplified Payment Verification (SPV), eliminates the need to download and verify the entire blockchain but, when implemented incorrectly, can create privacy risks~\cite{SPVbugs}.}

One subtlety of Bitcoin's transaction architecture is that in order to spend Bitcoins, the entire value of unspent outputs (\ie from previous transactions) must be spent. To accommodate this, Bitcoin clients automatically spend the full amount of unspent outputs and create multiple components in the transaction: one component will send part of the unspent coins to the intended recipient, and the other component will send the remaining inputs back to the sender as \emph{change}. It is technically possible (and some clients behave this way) to send change back to the sending address. However, to enhance anonymity, the reference client generates fresh addresses (and corresponding private keys) to receive the remaining transaction amount. 

As more transactions are made, Bitcoin clients must keep track of multiple private keys for use in future transactions. Many clients prominently display a Bitcoin balance on the main screen, which represents the sum of all unspent outputs for which private keys are available. 

\subsection{Usability of Key Management}

Passwords remain the most common form of user authentication~\cite{Herley.2012}. Private key-based authentication is rarely used by non-experts, and is typically never used as the default configuration in applications which support this authentication method. Transport Layer Security (TLS) client-side certificates have failed to reached wide-spread deployment. Secure shell (SSH) uses passwords by default, and allows certificates.


Password managers, when configured to generate or store system-chosen random passwords, share at least one property of cryptographic keys: such passwords become something you \emph{have} instead of \emph{know}. However, if access to such a password is lost, online services generally offer account recovery mechanisms (\eg based on email). No such recovery mechanism exists for self-managed cryptographic keys.

The use of public key systems by non-experts that is closest to Bitcoin is arguably encrypted/authenticated email, in particular Pretty Good Privacy (PGP) and its open-source alternatives (\ie GPG and OpenPGP). Beginning with \emph{Why Johnny Can't Encrypt}~\cite{WT99}, the usability of public key technology has been well-studied from a usability perspective~\cite{GM05,GMSN+05,SBKH06,GFF06}. The findings of this literature are diverse but relevant observations include the following: (1) the metaphor and terminology behind public and private keys is confusing; (2) it is difficult to correctly obtain other users' public keys; (3) key migration between devices is difficult. This literature tends to focus primarily on encryption and not signatures, but we find some overlap to the work presented here. 



\section{Bitcoin Key Management Approaches} 
\label{sec:approaches}

Before turning to a detailed usability evaluation, we evaluate from a systems perspective each category of tool for managing Bitcoin private keys. We highlight security and deployability issues, and note relevant details of the Bitcoin protocol that create complexities and potential discrepancies with users' mental models.

\subsection{Keys in Local Storage}
\label{sec:localstorage}
One way in which Bitcoin software manages several private keys is by storing these keys on the device's local storage, typically in a file or database in a pre-configured file system path. When a new transaction is created, the Bitcoin client can read the keys and immediately (possibly without any further user input) broadcast the transaction over the network. The reference Bitcoin client (Bitcoin Core), as well as certain mobile wallets (\eg Android Bitcoin Wallet) use this approach, storing private keys in a file (referred to as a \emph{wallet}) inside the user's home or application directory. 

Storing keys in a locally accessible file has several advantages. First, there is no additional cognitive load on users, since only the software must access the file. Second, a practically unlimited number of keys can be stored on disk due to the small size of keys. Third, the Bitcoin software can automatically generate keys and create transactions without additional input or actions from the user. 

Storing keys locally also creates several threats, which the user must consider. For example, the file storing private keys can be read by any application with access to the user's application folder. Malware authors may be particularly interested in exploiting this key management approach, since access to the local file results in the adversary gaining immediate access to the victim's funds. One of the first examples of private key-stealing malware was discovered by Symantec in 2011~\cite{coinbit}, with many other similar malware examples following suit.

Users must be cautious to not inadvertently share their Bitcoin application folder (\eg through peer-to-peer file sharing networks, off-site backups or on a shared network drive). Physical theft, especially in the case of portable computers or smartphones must also be considered. Similar to the storage of other sensitive files, threats to digital preservation~\cite{BKM05} should be taken into account. Examples include general equipment failure due to natural disasters and electrical failures; acts of war; mistaken erasure (\eg formatting the wrong drive or deleting the wrong folder); bit rot (\ie undetected storage failure); and possibly others. If storing private keys for a long period of time (\eg a trust fund or long-term savings), users must also preserve a specification of the file format to ensure the keys can continue to be read.

The reference Bitcoin client pre-generates keys in a batch of 100 (these keys are known as the keypool). When a transaction is made, the next available key is selected from the keypool for receiving change. The keypool is then periodically refilled with a new batch of keys as necessary. This \emph{key churn} requires users to periodically create new backups of their key storage file to ensure that new keypool keys are stored.

\subsection{Password-protected (Encrypted) Wallets}
Certain Bitcoin clients allow a locally stored wallet file to be encrypted with a key derived from a user-chosen password or passphrase. 
Password-protected wallets appear to address only \emph{physical} theft of the underlying storage device, requiring brute-force of the password if the file containing private keys is stolen. Password protection seems less useful in the case of \emph{digital} theft; if malware can be installed on to the device storing the wallet, it is reasonable to assume a keystroke-logging module would be present, limiting or nullifying the benefits of the password protection.

Password-protected wallets share the advantages and disadvantages of non-encrypted wallets (see Section~\ref{sec:localstorage}), with a few subtle differences. Password-protected wallets trade recoverability and usability for the mitigation of physical theft. If the password is forgotten, users lose the balance of their password-protected wallet since no mechanism exists for recovery\footnote{Of course, exhaustive search of the password space is theoretically possible, and is available as a service: \url{http://www.walletrecoveryservices.com}}. For day-to-day use, users must unlock the wallet by entering their password when new transactions are made. 


Password-protected wallets may mislead the user to believe that the password itself provides access to their funds regardless of the location of the device storing the wallet, as would be congruent with a traditional mental model for web-based online banking. Users may be surprised to discover that they cannot access their funds at a new device by simply entering their encryption password; the wallet file must also be transferred to the new device. 

\subsection{Offline Storage of Keys}
\label{sec:offline storage}
To further protect Bitcoin private keys from malware-based threats, wallets can be stored offline on some form of portable media, such as a USB thumbdrive. Keeping keys offline enables the use of traditional physical security techniques (\eg storing the drive in a fire-proof safe) to protect the wallet. However, offline storage has the drawback of making the wallet inaccessible for immediate use by software, preventing users from spending funds unless the offline storage media is nearby. As expected, offline storage can be used for backup, but all copies of the wallet must be kept offline for the full benefits of theft-protection to be realized. Prior to offline storage (wallet creation) and after storage (future transactions), the wallet will be exposed on a computational device, potentially to malware. 

\begin{figure}[pt]
  \includegraphics[width=.48\textwidth]{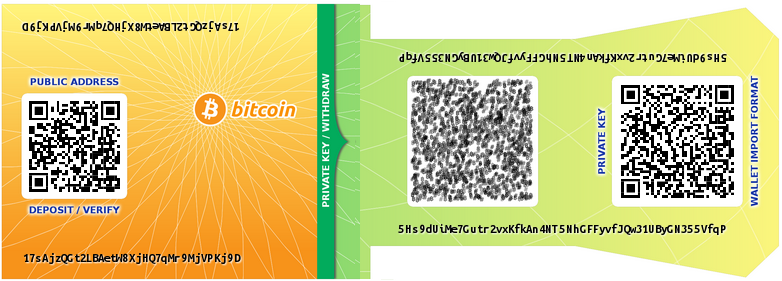}
  \caption{Bitcoin paper wallet generated using \url{https://bitcoinpaperwallet.com}. The printout is designed to be folded such that the private key (right) remains hidden while the public component (left) remains visible.}
  \label{fig:paperwallet}
\end{figure}

An interesting case of offline key storage is paper wallets (see ~\autoref{fig:paperwallet})
where private keys are printed onto paper typically in the form of a 2D barcode (\eg a QR code) or as a long sequence of characters. Barcodes facilitate reading the key back into a Bitcoin client by, for example, scanning the code with a smartphone camera. Securing a paper wallet is similar to securing cash, which most users should be comfortable with. However, funds can be stolen from a paper wallet by simply observing the QR code (\eg on live television\footnote{``A Bloomberg TV Host Gifted Bitcoin On Air And It Immediately Got Stolen,'' \textit{Business Insider}, 10/23/2013.}), which is not possible with physical money. Thus transporting a paper wallet securely requires that the printed contents remain unobservable at all times.  Users must remember that a paper wallet does not contain the funds itself, but rather enables signing authority over a set of Bitcoins. For example, if a paper wallet is discarded after funds are spent, the paper wallet still provides access to any future funds that may be sent to that address.\footnote{``Five Ways to Lose Money with Bitcoin Change Addresses,'' \textit{Bitzuma} (Blog), 17/03/2014.} 


As with any long-term storage, users must preserve software capable of decoding the QR code in the event that the paper wallet generation service is unavailable when attempting to reload keys onto a device. As of writing, many Bitcoin clients as well as offline storage solution use a common ``wallet import format'', which involves manipulating an ECDSA private key by performing cryptographic hashes, adding a checksum for integrity, and encoding the resulting string into Base58.\footnote{Base58 avoids the use of characters such as ``0, O, I, and l'' which may look visually similar, and also avoids punctuation characters which may trigger software (\eg e-mail clients) to perform line breaks.}

\subsection{Air-gapped Key Storage}
In offline storage, we assume the device or media holding private keys cannot perform computations such as creating digital signatures. We distinguish this type of storage from air-gapped storage, where wallets are stored on a secondary device that generates, signs, and exports transactions, but this secondary device is never connected to a network. When spending Bitcoins using an air-gapped device, a transaction is created from the air-gapped device and the resulting signed output transported (usually through portable media) to an Internet-enabled device for transmission onto the Bitcoin network. 

An air gap improves theft-resistance by never directly using a private key on an Internet-connected device. However, air gapped devices are capable of actually executing malware if infected. Malware may jump the air gap by infecting the portable media used to export signed transactions.

While not literally an air gap, hardware security modules (HSMs) emulate the properties of an air gap by isolating the key material from the host device, and only exposing the ability to sign transactions. Bitcoin-specific HSMs are under active development at the time of writing and a few have been recently released (\eg Trezor\footnote{\url{http://www.bitcointrezor.com}}).

Note that the consequences of obtaining access to the private keys are not much different from accessing a transaction-signing oracle for the wallet---both allow the current balance of Bitcoin to be stolen. However, future funds may be protected if access to the signing oracle is non-persistent. 

\subsection{Password-derived Keys} 
Thus far, all key management solutions have required users to maintain cryptographic keys. The remaining two solutions enable users to access their Bitcoin with a password instead. 

The first approach is to derive cryptographic keys from a user-chosen password (\eg using PBKDF2~\cite{pbkdf2}, manipulating the output to produce a valid Bitcoin private key). The disadvantage of using this approach directly is that only one resulting keypair is created, requiring the user to select a new (different) password for a new keypair. 

A more robust approach is described in the Bitcoin Improvement Proposal 32~\cite{bip32}, and is known as a Hierarchical Deterministic (HD) Wallet. HD wallets deterministically derive a set of private keys from a master secret (a randomly chosen passphrase). These keys can derive new private keys. The deterministic nature allows the password holder to view the balance, as well as spend the funds, of any sub-account derived from the password. However, if the private key on one of the sub-accounts is compromised, only the funds sent to that sub-key (or sub-keys derived from it) may be stolen. 


Password-derived wallets are targeted at loss-prevention and simpler cross-device access. The challenges of preserving access to a digital file are no longer necessary as long as the wallet can be re-generated from a memorized password. The primary drawback of a password-derived wallet is that weak user-chosen passwords can be found through unthrottled exhaustive search since a fingerprint of the associated public key will be in the global public ledger if the account holds any amount of Bitcoin. Rainbow tables~\cite{oechslin2003making} for password-derived keys have been developed.\footnote{D. Martyn. ``Bitcoin `Brainwallets' and why they are a bad idea,'' \textit{Insecurety Research \emph{(sic)}} (Blog), 26 Mar 2013.} Finally, it remains unclear whether memorization poses an advantage over maintaining a digital file when preventing loss---a forgotten password will orphan all funds in the account.

\subsection{Hosted Wallets} 
A final approach to key management is to host user accounts on a third-party web service. In this case, the service maintains possession of the private keys. Hosted wallet web services provide the user with access to transactional functionalities through standard web authentication mechanisms, such as a password or two-factor authentication, and may also offer password recovery mechanisms. Bitcoin smartphone applications that act as clients to hosted wallets benefit from reduced application complexity (\ie no need to perform cryptographic operations on the device) and brick and mortar bank-like user interfaces. Currency exchange services that allow Bitcoin to be exchanged with fiat currency effectively provide this service, as do web services deployed specifically to host wallets.


It is natural to expect hosted wallet services will become primary targets of attack since these services typically hold large amounts of Bitcoin. Offloading the task of key management to a third-party requires users to assume the risk that the service could be breached and funds lost, in exchange for a traditional online banking-style user experience. 

As a counter-measure to theft, hosted wallet providers often keep only a small float of their holdings online (called \textit{hot storage}) and store the majority of their holdings offline in \textit{cold storage}. This has the drawback of causing delays in transactions for users if the hot storage amount is exhausted.
Hosted wallet services may also allow audits, where they cryptographically prove possession of sufficient Bitcoin to match their liabilities.  

Another approach that falls under the hosted wallet category is a hybrid hosted wallet. Hybrid wallets use client side encryption (typically in Javascript) to encrypt all private keys and sensitive data. The web service is then only used for broadcasting transactions to the network and for displaying the user's balance (which requires inspecting the entire blockchain).




\begin{table*}[ht!]

\renewcommand{\arraystretch}{1.3}

\centering

\begin{tabular*}{0.9\textwidth}{@{\extracolsep{\fill}} llccccccccccccc}

\textit{Category} &
\textit{Example} & 
\headrow{Malware Resistant} & 
\headrow{Key(s) Kept Offline} &  
\headrow{No Trusted Third Party} &
\headrow{Resistant to Physical Theft} &
\headrow{Resistant to Physical Observation} &
\headrow{Resilient to Password Loss} & 
\headrow{Resilient to Key Churn} &
\headrow{Immediate Access to Funds} &
\headrow{No New User Software} & 
\headrow{Cross-device Portability} & 
\headrow{ } & 
\headrow{ } \\ \hline 

Keys in Local Storage & \bitcoinclient	&	&	&\full	&	&\full	&\full	&\full	&\full	&	&	&&\\
Password-protected Wallets &\multibit	&	&\prt	&\full	&\prt	&\full	&	&\full	&\full	&	&	&&\\
Offline Storage	&\paper				&\prt	&\full	&\full	&	&	&\full	&	&	&	&\full	&&\\ 
Air-gapped Storage & \armory 		&\prt	&\full	&\full	&	&\full	&\full	&\full	&	&	&	&&\\
Password-derived Keys & \brain		&	&\full	&\full	&\prt	&	&	&\full	&\full	&\full	&\full	&&\\ 
Hosted Wallet (Hot) & \coinbase			&	&	&	&	&	&\full	&\full	&\full	&\full	&\full	&&\\ 
Hosted Wallet (Cold)	&				&\prt	&\full	&	&	&	&\full	&\full	&	&\full	&\full	&&\\
Hosted Wallet (Hybrid)	&	\block			&	&\prt	&\prt	&	&	&\full	&\full	&\full	&\full	&\full	&&\\\hline
Cash &							&\full	&\full	&\full	&	&\full	&\full	&\full	&\full	&\full	&\full	&&\\ 
Online Banking &					&	&	&	&	&	&\full	&\full	&\full	&\full	&\full	&&\\ \hline 
\\
																					
\end{tabular*}

\caption{A comparison of key management techniques for Bitcoin (contrasted with traditional financial services). \full~ indicates the category of client is awarded the benefit in the corresponding column. \prt~partially awards the benefit. Details provided inline.}
\label{tab:prims}
\end{table*}

\section{Evaluation Framework}
In this section, we systematize the major category-wide issues we have uncovered in describing the various key management approaches used by Bitcoin clients. We present an evaluation framework based on 10 criteria as shown in Table~\ref{tab:prims} and discussed in the following subsections. This framework both summarizes the advantages and disadvantages of the various approaches we have evaluated, while also providing a benchmark for evaluating future key management proposals. The framework is adapted from a similar framework for evaluating password replacement schemes~\cite{BHOS12}.

\subsection{Evaluation Criteria}
We briefly enumerate the criteria used to evaluate each proposal in the framework below. 

\paragraph{Malware Resistant}
\label{Malware Resistant}
Malware designed to steal Bitcoin wallets and related passwords has been observed in the wild. Wallets that are not stored on an Internet-connected device, or devices capable of performing computations are considered malware resistant (\full), unless creating a transaction involves transferring to a computational device (\prt). 

\paragraph{Key Stored Offline}
\label{Key Kept Offline}
For archival storage of infrequently used keys, keys not directly accessible from an Internet-connected device---either due to being offline (\full) or online but password-protected (\prt)---are preferable. 

\paragraph{No Trusted Third Party}
\label{No Trusted Third Party}
All Bitcoin key management tools are trusted to a certain extent. 
This criteria considers the absence of a persistent trusted third party (\full) that maintains direct signing authority over a user's Bitcoin. 

\paragraph{Resistant to Physical Theft}
\label{Resistant to Physical Theft}
If the cryptographic keys are stored on some media or device that can be physically stolen, we do not consider the tool to be resistant to physical theft. Within our framework, the only tools meeting this requirement rely on a human memorized password being necessary for key recovery. These are awarded (\prt) since passwords tend to be weak and may not adequately resist unthrottled guessing. 

\paragraph{Resistant to Physical Observation}
\label{Resistant to Physical Observation}
Physical observation, such as observing key strokes or capturing QR codes with a camera, may result in access to a user's Bitcoin account. 

\paragraph{Resilient to Password Loss}
\label{Resilient to Password Loss}
If passwords are used (\prt), the loss of a password could result in some Bitcoin becoming unrecoverable if it is a necessary authentication factor in obtaining access to the signing key. For solutions where funds are held by third parties, these entities could provide a password recovery/reset mechanism (\full).

\paragraph{Resilient to Key Churn}
\label{Compatible with Change Keys}
Assuming the client sends change from transactions to a newly created change addresses, a tool is resilient to key churn if it can maintain access to the funds even after exhausting the initial keypool (\full). Tools not awarded this benefit are not guaranteed to maintain persistent access to new change addresses, and any balance sent to these addresses may be lost.

\paragraph{Immediate Access}
\label{Immediate Access}
Key management mechanisms that maintain direct access to the wallet enable Bitcoin to be transacted immediately (\full). We award this benefit to techniques that require a user to enter a password. We omit the benefit for techniques that require data to be obtained from external storage medium or secondary device. 

\paragraph{No New User Software}
\label{No New Software}
Some approaches require users to install new software on their system, for which the user may not have suitable permission, or software may not be developed for their specific platform (\eg some mobile platforms). By contrast, some tools can be executed from widely available software such as any standards-compliant web browser (\full). 

\paragraph{Cross-Device Portability}
\label{Portable}
A key management technique is cross-device portable (\full) if it allows easy sharing of the a Bitcoin address across multiple devices with minimal configuration or usability issues due to complexities like key churn.

\subsection{Discussion}
\autoref{tab:prims} demonstrates that key management approaches provide varying levels of security and convenience, with no single approach being obviously superior to others. One possible takeaway from our evaluation and comparison is that users can benefit heavily by offloading key management to a trusted party (\eg hosted wallets). The lower right side of the chart focuses on usability properties that are already present in traditional financial services (\ie resilient to password loss, no new software, cross-device portability). These properties are difficult to obtain if users independently manage their keys through one of the local storage techniques. Of course, the disadvantage of trusting a third party is that Bitcoin funds are now bound by a contractual agreement between users and the hosted wallet provider, negating one of the primary features of Bitcoin: a fully decentralized currency. Users in countries lacking regulatory maturity for digital currencies should exercise caution when trusting a third party with large amounts of Bitcoin. 

Based on our analysis, users can be given the concrete advice of treating digital currency much like they would treat fiat currency: keeping small amounts in ready-to-spend form (\eg local storage or online hosted walled) mimicking cash, and keeping larger sums in more difficult to access but more secure storage (\eg air-gapped or offline storage) mimicking a savings account or trust fund. Barber \etal~\cite{Barber2012} suggest the use of ``super wallets'' where users essentially run their own personal bank. A super-wallet keeps keys across multiple devices and requires all (or a subset using a threshold scheme) to be present to transfer funds to sub-wallets. Pre-configured transfers of small amounts can be authorized to move funds to sub-wallets that can be used for day-to-day spending. While the idea of super-wallets is intuitive, the implementation of such a scheme could introduce high levels of complexity.



\section{Usability Evaluation of Bitcoin Clients}

\subsection{Methodology}

We used a series of cognitive walkthroughs~\cite{WRLP94} to evaluate the usability of six Bitcoin clients. Cognitive walkthrough is a form of expert evaluation where an expert (or group of experts) steps through the design to evaluate aspects of its usability. The focus of the walkthrough is on the novice user and emphasizes \emph{learnability}. At each step, the evaluators ask three questions: Will the user see what to do? Will the user see how to do it? And once it is done, will the user know if they have performed the correct action?

We chose to use cognitive walkthroughs for several reasons. First, it allowed us to choose and compare standard tasks on disparate tools, and gave us easily compared insight into the common problems and successes of different Bitcoin clients. The cognitive walkthrough also allowed us to keep the focus on the novice user. The goal of our evaluation was to uncover problems specific to key management within Bitcoin software rather than to evaluate the usability of the clients themselves.



For our cognitive walkthrough, we defined a set of core tasks involving key management that a typical user needs to perform. We compared the results of each walkthrough against a standard set of evaluation guidelines, combining aspects of an heuristic evaluation~\cite{HeuristicEvaluation} with the walkthrough in order to interpret our results. 

Each of the following four tasks was independently performed by 2 experts to evaluate each tool:

\begin{compactlist}
	\item[\bf T1] Configure a new Bitcoin address and obtain its balance. This task involves launching the Bitcoin client (or logging into one if hosted online) for the first time. After a new address has been generated (either explicitly or transparently in the background), the user should be confident that the address' balance is XBT~0.00000000. The user should also be able to find their receiving Bitcoin address.\label{sec:ct-1}
	\item[\bf T2] Spend Bitcoin. Send some amount of Bitcoin to an arbitrary (but valid) Bitcoin address. This task requires the user to create a new transaction, entering relevant information such as recipient, amount, \etc\label{sec:ct-2}
	\item[\bf T3] Spend Bitcoin from the same address as above, but on a secondary device. This task may require copying private keys to the secondary device, entering passwords on multiple devices, or logging in to a hosted wallet provider on a different browser.\label{sec:ct-3}
	\item[\bf T4] Recover from the loss of the main credential. In the case of locally stored keys, this task involves restoring a file from backup. Otherwise this task involves recovering from password loss.\label{sec:ct-4}
\end{compactlist}

Since the focus of our walkthrough was on configuration and learnability, we used a set of heuristics first developed for a usability evaluation of Tor~\cite{COA07}. We chose to use these guidelines because like the anonymity software, successfully managing Bitcoin involves the application of complex cryptographic knowledge in an everyday activity. The set of guidelines, from~\cite{COA07}, are: 


\begin{compactlist}
	\item[\bf G1] Users should be aware of the steps they have to perform to complete a core task.
	\item[\bf G2] Users should be able to determine how to perform these steps.
	\item[\bf G3] Users should know when they have successfully completed a core task.
	\item[\bf G4] Users should be able to recognize, diagnose, and recover from non-critical errors.
	\item[\bf G5] Users should not make dangerous errors from which they cannot recover.
	\item[\bf G6] Users should be comfortable with the terminology used in any interface dialogues or documentation.
	\item[\bf G7] Users should be sufficiently comfortable with the interface to continue using it.
	\item[\bf G8] Users should be aware of the application's status at all times.
\end{compactlist}


\subsection{Evaluated Clients}
Real-world evaluation of the general approaches detailed in \autoref{sec:approaches} is difficult. Thus, we select six distinct Bitcoin clients or utilities that implement the key management approaches described. For the purposes of our usability evaluation, each client was evaluated in its default configuration on OS X unless otherwise stated. 

\paragraph{Keys in Local Storage} The reference Bitcoin client, Bitcoin Core~\cite{bitcoinqt}, is a cross-platform client that stores keys locally (optionally encrypted with a password). Bitcoin Core is the first recommended client on the \url{bitcoin.org} website. 

\paragraph{Password-protected (Encrypted) Wallet} We use the MultiBit~\cite{multibit} client (also recommended on \url{bitcoin.org}) since it provides a more convenient way to encrypt with a user-chosen password.

\paragraph{Offline Storage} We use paper wallets as offline storage. While paper wallets can be as simple as printing private keys on to paper, we select the paper wallet creation website \url{Bitaddress.org}~\cite{bitaddress}. Bitaddress allows users to generate new randomized keys in their web browsers, and then print QR encoded keys. 

\paragraph{Air-gapped Storage} We select the Bitcoin Armory~\cite{bitcoinarmory} client which includes functionality for creating an offline wallet that can be used to sign and export transactions. 

\paragraph{Password-derived Keys} One of the simplest ways to create a password-derived key is on the Brainwallet~\cite{brainwallet} website. The site allows users to enter a passphrase which is converted into a private key. 

\paragraph{Hosted Wallets} We use \url{Blockchain.info}~\cite{blockchain} as our hosted wallet provider. As of writing, Blockchain.info advertises the management of over 2.5 million user wallets.



\section{Results}
For space reasons, we summarize the results of each task for each client. Detailed walkthroughs can be found in our full technical report.\footnote{\url{http://users.encs.concordia.ca/~clark/papers/2015_usec_full.pdf}}

\subsection{Keys in Local Storage (\bitcoinclient)}
\paragraph{T1: Configure} \bitcoinclient transparently generates a new set of addresses on first run, but shows no notification to the user that this has occurred (fails G3). The receiving address can be found under the \emph{Receive coins} tab, but this could be easily confused with the \emph{Addresses} tab which contains a contact list of other user addresses (fails G2). 

To retrieve the account balance, \bitcoinclient must be online and the user must wait until a full copy of the blockchain has been downloaded. Due to the size of the blockchain, this can take days to complete, which may lead users to think that the client is dysfunctional. A status bar displaying ``Synchronizing with network'' shows the progress of the blockchain download (achieves G8), but the terminology may be too technical for novice users (fails G4 and G6). Once the blockchain has been downloaded, the balance is displayed on the \emph{Overview} tab (achieves G3). 

\paragraph{T2: Spend} Spending Bitcoin is straightforward since the keys are readily available to the \bitcoinclient client. Users spend Bitcoin by navigating to the \emph{Spend} tab (achieves G1 and G2). Since our focus is on key management, we do not evaluate the actual completion of transactions (which may have additional usability issues). We focus on ensuring the key is available to the software tool (which is not so straightforward with \eg offline storage). 

\paragraph{T3: Spend from Secondary Device} Installing \bitcoinclient on a secondary device creates a new set of keys. Users may not understand that the keys must be copied to the secondary device (fails G1), and if so, what file must be copied (fails G2). No information is provided regarding the potential threats of transferring the key through an insecure channel (fails G5). After transferring the key file, there is no user facing menu to import the file (fails G2). The only way to import keys is to use an advanced debug menu, or to overwrite the key file on disk (fails G6 and G7), then a blockchain re-scan might be needed in order to show the correct balance (fails G3)

\paragraph{T4: Recovery} If only one device is used, there is no way to recover from loss of the key file (\eg due to a disk failure, file corruption, or loss of the device itself; fails G5). If the user backed-up the key file, the process for recovering from loss is equivalent to that of T3 above. 


\subsection{Password-protected Wallets (\multibit)}
\paragraph{T1: Configure} On first run, a welcome page contains an explanation of common tasks that can be performed with \multibit---where the send, request and transaction tabs are and how to password protect the wallet file (achieves G1 and G2). The client provides help options for other functionalities with direct and non-technical guides (achieves G6). 

\multibit automatically generates a new receiving address on first run, but does not notify the user (fails G3). Reading the newly generated address requires navigation to the \emph{Request} tab, which displays ``Your address'' (partially achieves G2).

\paragraph{T2: Spend} The user must navigate to the tab labeled \emph{Send}, as instructed on the welcome screen (achieves G1 and G2). If the client is not synced, the send button is disabled (achieves G4). when synced, the user fills out the destination address and amount and clicks send. The client prompts the user for the decryption password (achieves G2). An incorrect password displays the error `The wallet password is incorrect' but otherwise allows immediate and unlimited additional attempts. Entering the correct password authorizes the transaction (achieves G3).


\paragraph{T3: Spend from Secondary Device} On the primary device, the user must navigate to the \emph{Options} menu, and select \emph{Export private keys} under tools (fails G1 and G2). The interface displays a wizard requesting an export password as well as a file system path for the exported file to be saved. On the secondary device, the user must select \emph{Import private keys} from the \emph{Options} menu. After selecting the previously exported file, the wizard confirms the completion of the import (achieves G4) and the balance is updated to reflect the newly imported keys. The user can then create a new transaction as in T2.


\paragraph{T4: Recovery}.
As with \bitcoinclient, recovery is not possible if no backup of the wallet file was made. Creating a backup and importing it follows the same procedure as T3. Both the password and the backed up wallet are necessary for recovery.

\subsection{Air-gapped Key Storage (\armory)}
\label{air gap}
\paragraph{T1: Configure} On first run, \armory displays a welcome page offering the option to `Import Existing Wallet' and `Create Your First Wallet!' (achieves G1 and G2). Passphrase-protection is mandatory in \armory, and the user is warned about the importance of not forgetting the passphrase (achieves G5). \armory then displays a wizard to create a paper wallet or create a digital backup of the wallet (achieves G6). The user can then click on \emph{Receive Bitcoins} to display the Bitcoin address as well as balance (achieves G3 and G4). \armory is dependent on \bitcoinclient to show the account's balance from the blockchain. While the blockchain is being downloaded, status is displayed (achieves G6 and G8). A message is shown when the blockchain download has completed (achieves G6). 

\paragraph{T2 \& T3: Spend}
The distinction between primary and secondary devices is less clear given that the basic setup itself includes two devices: one online and one offline, but authorization of transactions uses the offline device. To authorize a transaction, the user may begin from \armory on the online or offline device (may not fully achieve G2). On either device, the user should click on `Offline Transactions' in the main window which displays a very detailed description of the steps involved (achieves G1, G2, and G6). On the online computer, the user clicks the option: \emph{Create New Offline Transaction}. The user will be asked to enter the transaction details to generate an unsigned transaction as a file to be transferred to the offline computer. As mentioned in this step's documentation, the unsigned transaction data has no private data (the exact data will ultimately be added to the public blockchain) and no harm can be done by an attacker who captures this file (achieves G5) other than learning the transaction is being prepared.

On the offline computer, the user clicks on \textit{Offline Transactions} and then \emph{Sign Offline Transaction} which prompts the user for the unsigned transaction data file. \armory asks the user to review all the transaction information, such as the amount and the receiving addresses (achieves G5). By clicking on the \textit{sign} button signed transaction data can be saved to a file. Text at the top of the window describes the current state of the file (signed) and what must be done (move to online device) to complete the transaction (achieves G1 and G2).

The signed file should be transferred to the online computer and be loaded through the same offline transaction window. When a signed transaction is detected, the \emph{Broadcast} button becomes clickable. By clicking on broadcast, the user can once more review transaction details, and receive confirmation that the Bitcoins have been sent (achieves G3 and G8).

\paragraph{T4: Recovery} Like \bitcoinclient and \multibit, \armory requires a backup of the wallet to be made, and without it, recovery is impossible. \armory encourages backups at many stages (achieves G1 and G2). Importing a backup is straightforward by using the \emph{Import or restore wallet} menu and then following the wizard depending on what type of backup was made (paper or digital). The restore wallet menu also allows users to verify the integrity of backups to ensure that files were correctly backed up or printouts have no inconsistencies. 

We note that despite scoring positively on most guidelines during the execution of tasks in air-gapped approach, \armory supports a multitude of features (\eg Message signing, Offline transactions) which novices would typically not need to use. These features, along with their corresponding menus and descriptions, may be the source of confusion while performing simpler tasks, However, evaluating this aspect of \armory and other clients was beyond our scope. 
\subsection{Offline Storage (\paper)}

\paragraph{T1: Configure} Upon visiting the \url{bitaddress.org}, the user is asked to move the mouse or enter random characters in a text box to generate a high-entropy random seed to be used to generate a private key associated with the Bitcoin address (achieves G1 and G2). Once enough entropy has been collected, the site redirects the user to a page that shows the Bitcoin key pair(achieves G3). The public key (Bitcoin address) is labeled \emph{Share} in green and the private key \emph{Secret} in red (helping achieve G5). The web site documentation provides link to 2 different services to obtain the balance of the newly generated address (achieves G2). In general \paper uses non-expert terminology and simple instructions (achieves G6).

\paragraph{T2 \& T3: Spend} Since the keys are printed on paper, there is no difference between authorizing from a primary or secondary device so we collapse the analysis of core tasks 2 and 3.
 
To send funds from a Bitcoin address that has been stored on a paper wallet, the user must import the private key into one of the wallet clients such as \armory or \block hosted wallet. Importing the private key may require scanning the QR code or typing in the private key, depending on the client. Once the key has been imported, the user can proceed to spend Bitcoins by the particular client. We note that if the client returns change to newly generated addresses, and the user does not spend the full amount on the paper wallet, subsequently importing the paper wallet onto a different device will likely display no funds (partially fails G5). 

\paragraph{T4: Recovery} Loss of a paper wallet makes the funds unrecoverable (fails G5). \paper prompts the user to acknowledge this fact (also mentioned in its short documentation) when creating a paper wallet (achieves G1).

\subsection{Password-Derived Keys (\brain)} 

\paragraph{T1: Configure} The \brain website displays by default a pre-generated address corresponding to an empty passphrase. The passphrase input field displays ``Long original sentence that does not appear in any song or literature. Never use empty passphrase. (SHA256)'', but no corresponding documentation explains the purpose of the passphrase or how it relates to the generated key (fails G1, G2, G6). Users may not notice that generation of keys is happening dynamically as they type in the characters, possibly preventing the user from noticing that the task is complete (fails G3). Once the address has been generated, retrieving the balance of that address requires an external service, but no suggestions are provided on the site (fails G1 and G2). The interface displays a number of other fields (\eg additional encodings of the public key) which may not be meaningful to novice users (fails G6 and G7). 

\paragraph{T2 \& T3: Spend} Spending Bitcoins from a password-derived wallet requires the user to import the private key into another client. The user should experience similar usability challenges as those detailed in the Offline Storage client above.

\paragraph{T4: Recovery} Forgetting the password of a password-derived key leads to funds becoming unrecoverable (fails G5). Users will typically return to the same website (\ie the \brain website) to extract private keys, but this may not be possible if the site is inaccessible (fails G5). 

\subsection{Hosted Wallets (\block)}
\label{hosted}

\paragraph{T1: Configure} The user navigates to the \block site and creates a new wallet by providing an email address and a (min) 10 character password (achieves G1 and G2). A message warning the user about the importance of not forgetting the password is displayed during registration (achieves G5). Next, a \emph{Wallet Recovery Mnemonic} is shown to the user as a backup in case the password is forgotten. The balance and address are immediately displayed (achieves G3). 

\paragraph{T2 \& T3: Spend}
\label{hosted transaction} 
Hosted wallets are accessible from any web browser, so creating transactions from many devices is straightforward. The user logs in to the site, clicks \emph{Send money} (achieves G1 and G2). After filling in the required fields, the user is informed that the Bitcoins have been sent (achieves G3). Some of \block 's error messages may be too technical for novice users. For example, \emph{No free outputs to spend} is displayed when transactions are created without sufficient funds (fails G6). 

\paragraph{T4: Recovery} To recover from a forgotten password, a wallet recovery mnemonic may be provided on the login page. By clicking on the \emph{Recover Wallet} button, the site will ask for the mnemonic phrase and the email address to send the new credentials (achieves G1 and G2). Another recovery option is to proactively make backups and import them in case recovery is needed. To do so, in the main wallet page, user has to click \emph{Import/Export} and exporting either an encrypted or unencrypted backup. 



\section{Discussion}





\subsection{Metaphors}

Bitcoin naturally invites a metaphor to traditional currency. This metaphor is often used in the clients (\eg send coins, receive coins, wallet), but does not always support their usability. The coin metaphor fails in both of the ways that user interface metaphors traditionally fail~\cite{metaphorpaper}: aspects of Bitcoin transactions do not easily fit the coin metaphor, and conversely, encourages users to overextend the metaphor. Both of these lead to confusion on the part of users.   

One way in which the metaphor of physical coins fails is in the sending and receiving of Bitcoin. In the physical world, the same physical token is almost always used to represent the same unit of currency (\ie giving money to a friend involves handing them the coin). However, when Bitcoins are exchanged, the private key is not transferred along with the balance. Private keys remain in possession of the sender, and can be reused and associated with new coins at a later time. 

Many of the evaluated clients use the word ``Send'' to describe authorizing (digitally signing) a transaction, and private keys are not mentioned in any of the evaluated clients at the moment of transaction. It may appear counter-intuitive that this is a bad thing, but never mentioning the existence of keys may cause further confusion. The password-protected wallets, (\eg Multibit) require the user to input their password, but do not clarify the reason for the password.

Addresses are another metaphor that relate to the issue of transacting. The evaluated clients use the word ``Address'' to refer to the public key associated with a private key held by some user. This seems to be a relatively successful metaphor: it emphasizes the public nature of the public key, and also divorces the user's perception of a relationship between the public and private keys. To momentarily extend the metaphor, a user is accustomed to the idea that they will need to share their address in order to receive an item. However, the private key is more akin to the key to their mailbox, and a user would never think that they should share their mailbox key in order to receive mail to an address. 

Another pervasive metaphor in the evaluated clients is the Bitcoin ``wallet'', where the user's Bitcoins are stored. The wallet metaphor is deeply entrenched in the foundations of Bitcoin. The reference client, \emph{Bitcoin Core}, stores private keys in a file named wallet.dat and the \emph{MultiBit} client invites users to ``create your first wallet!'' on first launch. The hosted clients also use the metaphor; Blockchain.info prominently shows a Wallet tab, under which users are invited to ``Create My Free Wallet". The wallet metaphor is descriptive for users, but fails to encompass the complexity of a user's collection of keys. In reality, the Bitcoin wallet contains private keys, but the term wallet is used to describe both the file storing the private keys, and the main interface of Bitcoin clients (as in Blockchain.info). This main interface sometimes includes a variety of other information, such as transaction history, address book, currency exchange rates, \etc


\subsection{Abstractions}

Abstraction and automation are complex issues for security software. Often, security is too complex to be completely automated, and the problem cases are often punted to the user (\eg in the case of TLS certificates~\cite{SSLSobey}).  

On first run, all of the evaluated clients transparently generate keypairs without informing the user. This behaviour continues as new transactions are made, where clients generate new addresses with no user notification (\eg for receiving change). It is unclear how well this abstraction works: while users do not need to be burdened with the knowledge of each private key, there are still situations in which a user might need to manage those keys, and the abstraction prevents users from doing so. Recovery from key loss depends on the existence of an up-to-date backup. While backup sounds like a simple task, in many of the evaluated clients, it involves finding the right menu (MultiBit), or the right file (Bitcoin Core). Some clients do prompt the user to back up their wallets (\eg Bitcoin Armory), but with the private keys so completely abstracted away, users may not even understand what they are backing up, or why. Key churn, and the consequent need for semi-regular backups complicate the issue even farther. 

The abstractions made in Bitcoin clients are sometimes beneficial for users, such as in the case of displaying a user's balance. A user's Bitcoin balance is typically made up of many small amounts corresponding to many private keys. However, most of the evaluated clients abstract these balances into a single figure. This highlights a usability disadvantage of paper wallets -- the user must manage these multiple balances manually, and there is no method of seeing an aggregate balance when multiple paper wallets are in use.

\subsection{Technical Language and Content}
\begin{figure}[htb]
        \centering
        \subfloat[\bitcoinclient]{
        \includegraphics[width=.48\textwidth]{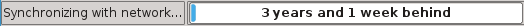}
        \label{fig:btcore-sync}
        }
                
        \subfloat[\multibit]{
        \includegraphics[width=.45\textwidth]{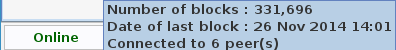}
        \label{fig:multibit-sync}
        }
        
        \caption{Screenshots of technical language displayed by two different clients. }\label{fig:techlang}
\end{figure}
When performing our evaluation, we identified multiple occurrences of highly specialized or technical language used in the Bitcoin clients. These instances of technical language are confusing, particularly to novice users who are unlikely to be aware of either the jargon, and for whom the language will not help clarify the issues. The language itself highlights the complexity of the tasks associated with Bitcoin, and the difficulty of explaining them simply.

Examples of such language included messages in \multibit and  \bitcoinclient that referenced the client being ``out of sync'' or ``synchronizing with network'' (see Figure~\ref{fig:btcore-sync}) referring the process of downloading a full copy of the blockchain or retrieving relevant blocks from a trusted peer. A related message in \multibit (Figure~\ref{fig:multibit-sync}) and \armory displayed the number of blocks that had been downloaded, as well as the number of connections to the Bitcoin network. These messages are intended to communicate that clients may benefit from faster transaction notifications when connected to more peers, but since peer connectivity is difficult for users to control, there is little benefit in communicating these ideas with the user. Similarly, the number of blocks independently has little significance to most tasks performed by an end user. We suggest that not only could this language be clarified, but that the interfaces could also streamline the amount of information that is presented to the user on every screen.

We also noticed that some clients used highly technical language when they could have used the metaphor to provide a simpler explanation to users. When attempting to authorize a new transaction on \block with insufficient funds, the web interface displayed ``no free outputs to spend''. This error message is confusing, and would be more easily understood if it referred to the lack of coins instead of the lack of outputs. Similarly, essential actions such as importing or exporting keys were often buried behind advanced or debug menus. 





In the evaluated clients, there were often few resources to which users could turn for help. In the cognitive walkthroughs, the answer to the question ``will the user know what to do?" was almost always unclear. Interface cues and features such as tool tips, wizards, or other contextual help were almost entirely lacking. Some actions were guided (\eg Multibit's prompted backups or create your first wallet), but many actions such as obtaining the balance of a paper or password-derived address were unsupported by help or documentation.


\section{Conclusion}
Bitcoin's usability limitations, particularly those related to key management, pose challenges to its rising popularity. In our evaluation, 
we found that developers in the Bitcoin ecosystem are making innovative attempts at solving the decades-old problem of usable key management. While some of these techniques seem promising, we find that tasks involving key management can be mired in complex metaphors and confusing abstractions.

Further investigation is needed to better understand and address these issues. A user study would give insight into exactly how these problems are affecting users and it would be interesting to investigate how expert users are (apparently successfully) handling these challenges. Bitcoin presents a new opportunity for public key cryptography to become mainstream, and our evaluation is a first step towards achieving usable key management in decentralized cryptocurrencies.

\bibliographystyle{IEEEtranS}
\footnotesize
\bibliography{IEEEabrv,bib/btc}
\normalsize


\end{document}